\documentclass[aps,nofootinbib,longbibliography,prd,twocolumn]{revtex4-1}
\usepackage[babel]{csquotes}
\usepackage{graphicx}
\usepackage{amsmath,amssymb}
\usepackage[colorlinks,citecolor=blue,linkcolor=blue,urlcolor=blue]{hyperref}
\usepackage{mathrsfs}
\usepackage{enumerate}
\def\be{\begin{equation}}
\def\ee{\end{equation}}

\begin{document}

\title{Meaning = Information + Evolution}
\author{Carlo Rovelli}
\affiliation{\small CPT, Aix-Marseille Universit\'e, Universit\'e de Toulon, CNRS, F-13288 Marseille, France.} 

\begin{abstract}
\noindent 
Notions like meaning, signal, intentionality, are difficult to relate to a physical word. I study a \emph{purely physical} definition of ``meaningful information", from which these notions can be derived. It is inspired by a model recently illustrated by Kolchinsky and Wolpert, and improves on Dretske classic work on the relation between knowledge and information. I discuss what makes a physical process into a ``signal". 
\end{abstract}

\maketitle

\section{Introduction}

There is a gap in our understanding of the world. On the one hand we have the physical universe; on the other, notions like meaning, intentionality, agency, purpose, function and similar, which we employ for the like of life, humans, the economy...    These notions are absent in elementary physics, and their  placement into a physicalist world view is delicate \cite{Price2011}, to the point that the existence of this gap is commonly presented as the strongest argument against naturalism.  

Two historical ideas have contributed tools to bridge the gap.  

The first is Darwin's theory, which offers evidence on how function and purpose can emerge from natural variability and natural selection of structures \cite{Darwin2009}.   Darwin's theory provides a naturalistic account for the ubiquitous presence of function and purpose in biology. It falls sort of bridging the gap between physics and meaning, or intentionality. 

The second is the notion of `information', which is increasingly capturing the attention of scientists and philosophers. Information has been pointed out as a key element of the link between the two sides of the gap, for instance in the classic work of Fred Dretske \cite{FredDretske1981}. 

However, the word `information' is highly ambiguous.  It is  used with a variety of distinct meanings, that cover a spectrum  ranging from mental and semantic (``the information stored in your USB flash drive is comprehensible") all the way down to strictly engineeristic (``the information stored in your USB flash drive is 32 Giga"). This ambiguity is a source of confusion.  In Dretske's book, information is introduced on the base of Shannon's theory \cite{Shannon:1948fk}, explicitly interpreted as a formal theory that ``does not say what information is". 

In this note, I make two observations.   The first is that it is possible to extract from the work of Shannon a \emph{purely physical}  version of the notion of information.  Shannon calls its ``relative information". I keep his terminology even if the ambiguity of these terms risks to lead to continue the misunderstanding; it would probably be better to call it simply `correlation', since this is what it ultimately is: downright crude physical correlation.

The second observation is that the combination of \emph{this} notion with Darwin's mechanism provides the ground for a definition of meaning.  More precisely, it provides the ground for the definition of a notion of ``meaningful information", a notion that on the one hand is solely built on physics, on the other can underpin intentionality, meaning, purpose, and is a key ingredient for agency. 

The claim here is not that the full content of what we call intentionality, meaning, purpose --say in human psychology, or  linguistics-- is nothing else than the \emph{meaningful information} defined here.  But it is that these notions can be built upon  the notion of \emph{meaningful information} step by step, adding the articulation proper to our neural, mental, linguistic, social, etcetera, complexity.  In other words, I am not claiming of giving here the full chain from physics to  mental, but rather the  crucial first link of the chain. 

The definition of meaningful information I give here is inspired by a simple model presented by David Wolpert and Artemy Kolchinsky \cite{Wolpert2016}, which I describe  below. The model illustrates how two physical notions, combined, give rise to a notion we usually ascribe to the non-physical side of the gap: meaningful information. 

The note is organised as follows.  I start by a careful formulation of the notion of correlation (Shannon's relative information).  I consider this a main motivation for this note: emphasise the commonly forgotten fact that such a purely physical definition of information exists.  I then briefly recall a couple of points regarding Darwinian evolution which are relevant here, and I introduce (one of the many possible) characterisation of living beings. I then describe Wolpert's model and give explicitly the definition of meaningful information which is the main purpose of this note.  Finally, I describe how this notion might bridge between the two sides of gap. I close with a discussion of the notion of signal and with some general considerations. 

\section{Relative information}

Consider physical systems $A, B, ... $ whose states are described by a physical variables $x, y, ...$, respectively.   This is the standard conceptual setting of physics.   For simplicity, say at first that the variables take only discrete values.  Let $N_{\scriptscriptstyle\! A}, N_{\scriptscriptstyle\! B}, ... $ be the number of distinct values that the variables  $x, y, ... $ can take.  If there is no relation or constraint between the systems $A$ and $B$, then the pair of system $(A,B)$ can be in $N_{\scriptscriptstyle\! A} \times N_{\scriptscriptstyle\! B}$ states, one for each choice of a value for each of the two variables $x$ and $y$. In physics, however, there are routinely constraints between systems that make certain states impossible. Let $N_{\scriptscriptstyle\! AB}$ be the number of allowed possibilities. Using this, we can define `relative information' as follows.

We say that $A$ and $B$ `have information about one another' if $N_{\scriptscriptstyle\! AB}$ is strictly smaller than the product $N_{\scriptscriptstyle\! A} \times N_{\scriptscriptstyle\! B}$. We call 
\be
S = \log (N_{\scriptscriptstyle\! A} \times N_{\scriptscriptstyle\! B}) - \log N_{\scriptscriptstyle\! AB}, 
\ee
where the logarithm is taken in base 2, the ``relative information" that $A$ and $B$ have about one another. The unit of information is called `bit'. 

For instance, each end of a magnetic compass can be either a North $(N)$ or South $(S)$ magnetic pole, but they cannot be both $N$ or both $S$. The number of possible states of each pole of the compass is 2 (either $N$ or $S$), so $N_{\scriptscriptstyle\! A}=N_{\scriptscriptstyle\! B}=2$, but the physically allowed possibilities are not  $N_{\scriptscriptstyle\! A} \times N_{\scriptscriptstyle\! B}=2\times 2=4$ $(NN,NS,SN,SS)$. Rather,  they are only two $(NS,SN)$, therefore $N_{\scriptscriptstyle\! AB}=2$. This is dictated by the physics. Then we say that the state ($N$ or $S$) of one end of the compass `has relative information'  
\be 
S = \log 2 + \log 2 - \log 2 = 1
\ee
 (that is: 1 bit) about the state of the other end.  Notice that this definition captures the physical underpinning to the fact that ``if we know the polarity of one pole of the compass then we also know (have information about)  the polarity of the other."  But the definition itself is completely physical, and makes no reference to semantics or subjectivity.
 
The generalisation to continuous variables is straightforward. Let $P_{\scriptscriptstyle\! A}$ and $P_{\scriptscriptstyle\! B}$ be the phase spaces of $A$ and $B$ respectively and let $P_{\scriptscriptstyle\! AB}$ be the subspace of the Cartesian product $P_{\scriptscriptstyle\! A} \times P_{\scriptscriptstyle\! B}$ which is allowed by the constraints. Then the relative information is 
\be
                                        S = \log V(P_{\scriptscriptstyle\! A} \times P_{\scriptscriptstyle\! B}) - \log V(P_{\scriptscriptstyle\! AB}) 
\ee
whenever this is defined.\footnote{Here $V(.)$ is the Liouville volume and the difference between the two volumes can be defined as the limit of a regularisation even when the two terms individually diverge. For instance, if $A$ and $B$ are both free particles on a  circle of of size $L$, constrained to be at a distance less of equal to $L/N$ (say by a rope tying them), then we can easily regularise the phase space volume by bounding the momenta, and we get $S = \log N$, independently from the regularisation.} 

Since the notion of relative information captures correlations, it extends very naturally to random variables. Two random variables $x$ and $y$ described by a probability distribution $p_{\scriptscriptstyle\! AB}(x,y)$ are uncorrelated if 
\be
                                             p_{\scriptscriptstyle\! AB}(x,y)=\tilde p_{\scriptscriptstyle\! AB}(x,y) 
\ee
where $\tilde p_{\scriptscriptstyle\! AB}(x,y) $ is called the marginalisation of $p_{\scriptscriptstyle\! AB}(x,y) $ and is defined as the product of the two marginal distributions 
\be
\tilde p_{\scriptscriptstyle\! AB}(x,y) = p_{\scriptscriptstyle\! A}(x)\, p_{\scriptscriptstyle\! B}(y), 
\ee
in turn defined by 
\be
p_{\scriptscriptstyle\! A}(x)=  \int p_{\scriptscriptstyle\! AB}(x,y)\ dy, \hspace{4mm} p_{\scriptscriptstyle\! B}(y)=  \int p_{\scriptscriptstyle\! AB}(x,y)\ dx.
\ee
Otherwise they are correlated. The amount of correlation is given by the difference between the entropies of the two distributions  $p_{\scriptscriptstyle\! A}(x,y)$ and $\tilde p_{\scriptscriptstyle\! A}(x,y)$. The entropy of a probability distribution $p$ being $S=\int p \log p$ on the relevant space. All integrals are taken with the Luoiville measures of the corresponding phase spaces. 

Correlations can exist because of physical laws or because of specific physical situations, or arrangements or mechanisms, or the past history of physical systems. 

Here are few examples.  The fact that the two poles of a magnet cannot have the same polarisation is excluded by one of the Maxwell equations.  It is just a fact of the world.  The fact that two particles tied by a rope cannot move apart more than the distance of the rope is a consequence of a direct mechanical constraint: the rope. The frequency of the light emitted by a hot piece of metal is correlated to the temperature of the metal at the moment of the emission. The direction of the photons emitted from an object is correlated to the position of the object. In this case emission is the mechanism that enforces the correlation.  The world teams with correlated quantities. Relative information is, accordingly, naturally ubiquitous. 

Precisely because it is purely physical and so ubiquitous,  relative information is not sufficient to account for meaning. `Meaning' must be grounded on something else, far more specific.

\section{Survival advantage and purpose}

Life is a characteristic phenomenon we observe on the surface of the Earth. It is largely formed by individual organisms that interact with their environment and embody mechanisms that keep themselves away from thermal equilibrium using available free energy.  A dead organism decays rapidly to thermal equilibrium, while an organism which is alive does not. I take this --with quite a degree of arbitrariness-- as a characteristic feature of organisms that are alive. 

The key of Darwin's discovery is that we can legitimately reverse the causal relation between the existence of the mechanism and its function.  The fact that the mechanism exhibits a purpose ---ultimately to maintain the organism alive and reproduce it--- can be simply understood as an indirect consequence, not a cause, of its existence and its structure. 

As Darwin points out in his book, the idea is ancient. It can be traced at least to Empedocles. Empedocles suggested that life on Earth may be the result of random happening of structures, all of which perish except those that happen to survive, and these are the living organisms.\footnote{[There could be] ``beings where it happens as if everything was organised in view of a purpose, while actually things have been  structured appropriately only by chance; and the things that happen not to be organised adequately, perished, as Empedocles says". \cite{Aristotle} II 8, 198b29)}   

The idea was criticised by Aristotle, on the ground that we see organisms being born with structures already suitable for survival, and not being born at random (\cite{Aristotle} II 8, 198b35). But shifted from the individual to the species, and coupled with the understanding of inheritance and, later, genetics, the idea has turned out to be correct. Darwin clarified the role of variability and selection in the evolution of structures and molecular biology illustrated how this may work in concrete. Function emerges naturally and the obvious purposes that living matter exhibits can be understood as a consequence of variability and selection.  What functions is there because it functions: hence it has survived. We do not need something external to the workings of nature to account for the appearance of function and purpose. 

But variability and selection alone may account for function and purpose, but are not sufficient to account for meaning, because meaning has semantic and intentional connotations that are not a priori necessary for variability and selection.  `Meaning' must be grounded on something else. 

\section{Kolchinsky-Wolpert's model and meaningful information}

My aim is now to distinguish the correlations that are are ubiquitous in nature from those that we count as relevant information.  To this aim, the key point is that surviving mechanisms survive by using correlations. This is how relevance is added to correlations. 

The life of an organisms progresses in a continuous exchange with the external environment. The mechanisms that lead to survival and reproduction are adapted by evolution to a certain environment.  But  in general environment is in constant variation, in a manner often poorly predictable. It  is obviously advantageous to be appropriately correlated with the external environment, because survival probability is maximised by adopting different behaviour in different environmental conditions.   

A bacterium that swims to the left when nutrients are on the left and swims to the right when nutrients are on the right prospers; a bacterium that swims at random has less chances. Therefore many bacteria we see around us are of the first kind, not of the second kind.  This simple observation leads to the Kolchinsky-Wolpert model  \cite{Wolpert2016},.

A living system $A$ is characterised by a number of variables $x_n$ that describe its structure. These may be numerous, but are in a far smaller number than those describing the full microphysics of $A$ (say, the exact position of each water molecule in a cell). Therefore the variables $x_n$ are macroscopic in the sense of statistical mechanics and there is an entropy $S(x_n)$ associated to them, which counts the number of the corresponding microstates.  As long as an organism is alive, $S(x_n)$ remains far lower than its thermal-equilibrium value $S_{max}$.  This capacity of keeping itself outside of thermal equilibrium,  utilising free energy, is a crucial aspects of systems that are alive.  Living organisms have generally a rather sharp distinction between their state of being alive or dead, and we can represent it as a threshold $S_{thr}$ in their entropy. 

Call $B$ the environment and let $y_n$ denote a set of variables specifying its state.   Incomplete specification of the state of the environment can be described in terms of probabilities, and therefore the evolution of the environment is itself predictable at best probabilistically.  

\begin{figure}
\includegraphics[height=4cm]{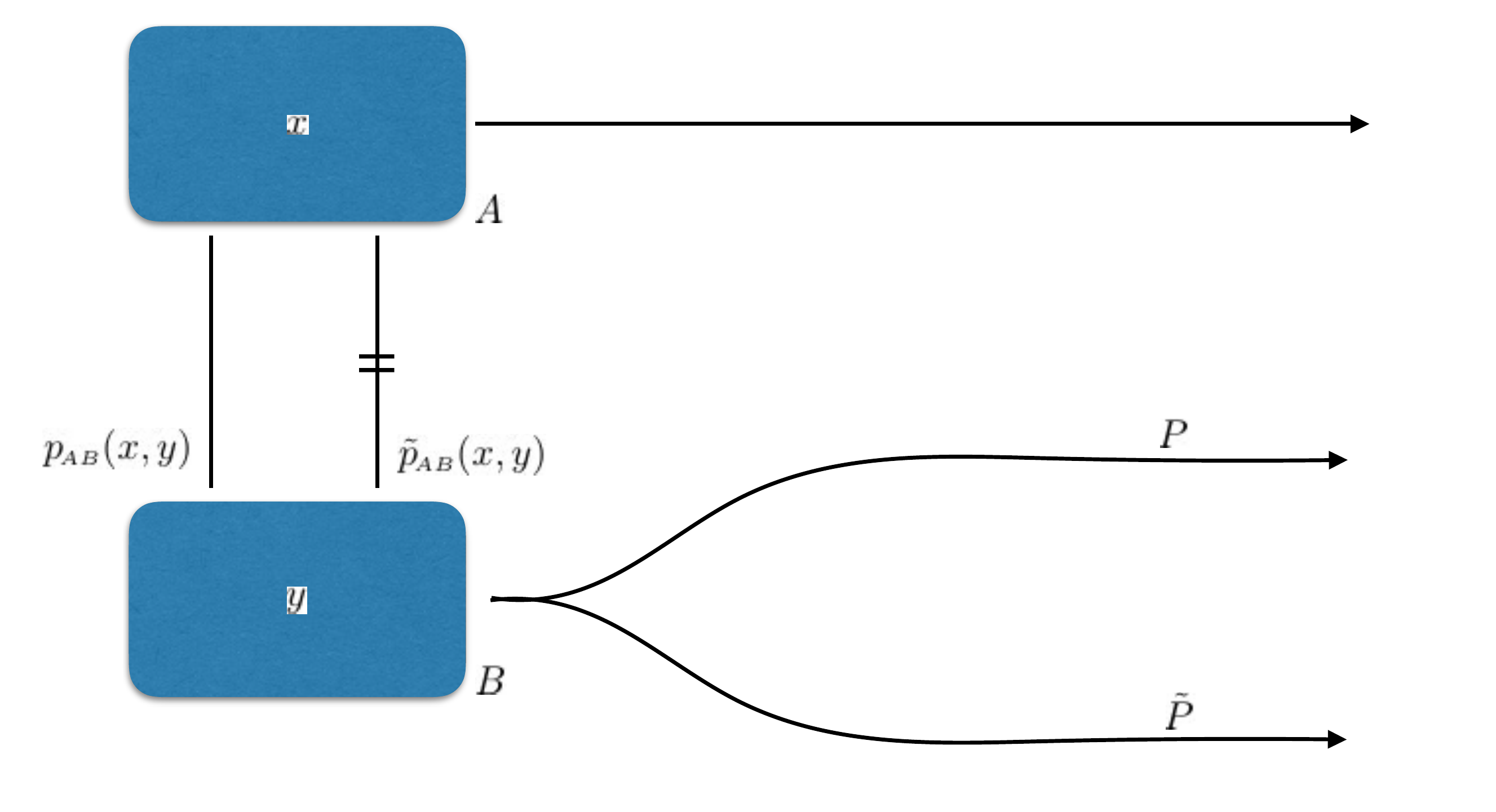}
\caption{The Kolchinsky-Wolpert model and the definition of meaningful information. If the probability of descending to thermal equilibrium $\tilde P$ increases when we cut the information link between $A$ and $B$, then the relative information (correlation) between the variables $x$ and $y$ is ``meaningful information".}
\label{WP}
\end{figure}

Consider now a specific variable $x$ of the system $A$ and a specific variable $y$ of the system $B$ in a given macroscopic  state of the world.   Given a value $(x,y)$, and taking into account the probabilistic nature of evolution, at a later time $t$ the system $A$ will find itself in a configuration $x_n$ with probability $p_{x,y}(x_n)$.  If at time zero $p(x,y)$ is the joint probability distribution of $x$ and $y$, the probability that at time $t$ the system $A$ will have entropy higher that the threshold is 
\be
         P=\int dx_n  dx\,dy\ p(x,y)\, p_{x,y}(x_n) \theta(S(x_n)-S_{thr}), 
\ee
where $\theta$ is the step function.  Let us now define 
\be
         \tilde P=\int dx_n  dx\,dy\ \tilde p(x,y)\, p_{x,y}(x_n) \theta(S(x_n)-S_{thr}).
\ee
where $\tilde p(x,y)$ is the marginalisation of $p(x,y)$ defined above. This is the probability of having above threshold entropy if we erase the relative information. This is Wolpert's model.

Let's define the relative information between $x$ and $y$ contained in $p(x,y)$ to be ``directly meaningful" for $B$ over the time span $t$, iff $\tilde P$ is different from $P$. And call 
\be
M=\tilde P-P
\ee
the ``significance" of this information. The significance of the information is its relevance for the survival, that is, its capacity of affecting the survival probability.   

Furthermore, call the relative information between $x$ and $y$ simply ``meaningful" if it is directly meaningful or if its marginalisation decreases the probability of acquiring information that can be meaningful, possibly in a different context. 

Here is an example. Let $B$ be food for a bacterium and $A$ the bacterium, in a situation of food shortage. Let $x$ be the location of the nurture, for simplicity say it can be either at the left of at the right. Let $y$ the variable that describe the internal state of the bacterium which determines the direction in which the bacterium will move. If the two variables $x$ and $y$ are correlated in the right manner, the bacterium reaches the food and its chances of survival are higher.  Therefore the correlation between $y$ and $x$ is ``directly meaningful" for the bacterium, according to the definition given, because marginalising $p(x,y)$, namely erasing the relative information increases the probability of starvation. 

Next, consider the same case, but in a situation of food abundance. In this case the correlation between $x$ and $y$ has no direct effect on the survival probability, because there is no risk of starvation. Therefore the $x$-$y$ correlation is not directly meaningful.  However, it is still (indirectly) meaningful, because it empowers the bacterium with a correlation that has a chance to affect its survival probability in another situation.\\ 

A few observations about this definition:
\begin{enumerate}[i.]

\item Intentionality is built into the definition. The information here is information that the system $A$ has about the variable $y$ of the system $B$. It is by definition information ``about something external".  It refers to a physical configuration of $A$ (namely the value of its variable $x$), insofar as this variables is correlated to something external (it `knows' something external).  

\item The definition separates correlations of two kinds: accidental correlations that are ubiquitous in nature and have no effect on living beings, no role in semantic, no use, and correlations that contribute to survival.
The notion of meaningful correlation captures the fact that information can have ``value" in a darwinian sense. The value is defined here a posteriori as the increase of survival chances. It is a ``value" only in the sense that it increases these chances. 

\item Obviously, not any manifestation of meaning, purpose, intentionality or value is \emph{directly} meaningful, according to the definition above. Reading today's newspaper is not likely to directly enhance mine or my gene's survival probability. This is the sense of the distinction between `direct' meaningful information and meaningful information. The second includes all relative information which in turn increases the probability of acquiring meaningful information.  This opens the door to recursive growth of meaningful information and arbitrary increase of semantic complexity.  It is this secondary recursive growth that grounds the use of meaningful information in the brain. Starting with meaningful information in the sense defined here, we get something that looks more and more like the full  notions of meaning we use in various contexts, by adding articulations and moving up to contexts where there is a brain, language, society, norms... 

\item A notion of `truth' of the information, or `veracity' of the information, is implicitly determined by the definition given.  To see this, consider the case of the bacterium and the food.  The variable $x$ of the bacterium can take to values, say $L$ and $R$, where $L$ is the variable conducting the bacterium to swim to the Right and $L$ to the Left.  Here the definition leads to the idea that $R$ \emph{means} ``food is on the right" and $L$  \emph{means} ``food is on the left".  The variable $x$ contains this information.   If for some reason the variable $x$ is on $L$ but the food happens to be on the Right, then the information contained in $x$ is ``not true".   This is a very indirect and in a sense deflationary notion of truth, based on the effectiveness of the consequence of holding something for true.  (Approximate coarse grained knowledge is still knowledge, to the extent it is somehow effective. To fine grain it, we need additional knowledge, which is more powerful because it is more effective.) Notice that this notion of truth is very close to the one common today in the natural sciences when we say that the `truth' of a theory is the success of its predictions. In fact, it is the same. 

\item The definition of `meaningful' considered here does not directly refer to anything mental.  To have something mental you need a mind and to have a mind you need a brain, and its rich capacity of elaborating and working with information.  The question addressed here is what is the physical base of the information that brains work with.  The answer suggested is that it is just physical correlation between internal and external variables affecting survival either directly or, potentially, indirectly. 

\end{enumerate}

The idea put forward is that what grounds all this is direct meaningful information, namely \emph{strictly physical} correlations between a living organism and the external environment that have survival and reproductive value.  The semantic notions of information and meaning are ultimately tied to their Darwinian evolutionary origin. The suggestion is that the notion of meaningful information serves as a ground for the foundation of meaning. That is, it could offer the link between the purely physical world and the world of meaning, purpose, intentionality and value. It could bridge the gap.

\section{Signals, reduction and modality}

A signal is a physical event that conveys meaning. A ring of my phone, for instance, is a signal that \emph{means} that somebody is calling. When I hear it, I understand its meaning and I may reach the phone and answer.

As a purely physical event, the ring happens to physically cause a cascade of physical events, such as the vibration of air molecules, complex firing of nerves in my brain, etcetera, which can in principle be described in terms of purely physical causation.   What distinguishes its being a signal, from its being a simple link in a physical causation chain? 

The question becomes particularly interesting in the context of biology and especially molecular biology.  Here the minute working of life is heavily described in terms of signals and information carriers: DNA codes the \emph{information} on the structure of the organism and in particular on the specific proteins that are going to be produced, RNA carries this \emph{information} outside the nucleus, receptors on the cell surface \emph{signal} relevant external condition by means of suitable chemical cascades. Similarly, the optical nerve exchanges \emph{information} between the eye and the brain, the immune system receives \emph{information} about infections, hormones \emph{signal} to organs that it is time to do this and that, and so on, at libitum. We describe the working of life in heavily informational terms at every level. What does this mean? In which sense are these processes distinct from purely physical processes to which we do not usually employ an informational language? 

I see only one possible answer. First, in all these processes the carrier of the information could be  somewhat easily replaced with something else without substantially altering the overall process.  The ring of my phone can be replaced by a beep, or a vibration. To decode its meaning is the process that recognises these alternatives as equivalent in some sense. We can easily imagine an alternative version of life where the meaning of two letters is swapped in the genetic code.  Second, in each of these cases the information carrier is physically correlated with something else (a protein, a condition outside the cell, a visual image in the eye, an infection, a phone call...) in such a way that breaking the correlation could damage the organism to some degree. This is precisely the definition of meaningful information studied here.

I close with two general considerations.

The first is about reductionism.  Reductionism is often overstated. Nature appears to be formed by a relative simple ensemble of elementary ingredients obeying relatively elementary laws. The possible combinations of these elements, however, are stupefying in number and variety, and largely outside the possibility that \emph{we} could compute or deduce them from nature's elementary ingredients. These combinations happen to form higher level structures that we can in part understand directly. These we call emergent. They have a level of autonomy from elementary physics in two senses: they can be studied  \emph{independently} from elementary physics, and they can be realized \emph{in different manners} from elementary constituents, so that their elementary constituents are in a sense irrelevant to our understanding of them. Because of this, it would obviously be useless and self defeating to try to replace all the study of nature with physics.  But evidence is strong that nature is unitary and coherent, and its manifestations are ---whether we understand them or not--- behaviour of an underlying physical world. Thus, we study thermal phenomena in terms of entropy, chemistry in terms of chemical affinity, biology in terms functions, psychology in terms of emotions and so on. But we increase our understanding of nature when we understand how the basic concept of a science are ground in physics, or are ground in a science which is ground on physics, as we have largely been able to do for chemical bonds or entropy. It is in this sense, and only in this sense, that I am suggesting that meaningful information could provide the link between different levels of our description of the world. 

The second consideration concerns the conceptual structure on which the definition of meaningful information proposed here is based.  The definition has a modal core.  Correlation is not defined in terms of how things \emph{are}, but in terms of how they \emph{could} or \emph{could not} be. Without this, the notion of correlation cannot be constructed. The fact that something is red and something else is red, does not count as a correlation.  What counts as a correlation is, say, if two things  \emph{can} each be of different colours, but the two \emph{must} always be of the same colour.  This requires modal language. If the world is what it is, where does modality comes from? 

The question is brought forward by the fact that the definition of meaning given here is  modal, but does not bear on whether this definition is genuinely physical or not.  The definition is genuinely physical.  It is physics itself which is heavily modal. 
Even without disturbing quantum theory or other aspects of modern physics, already the basic structures of classical mechanics are heavily modal. The phase space of a physical system is the list of the configurations in which the system \emph{can} be. Physics is not a science about how the world \emph{is}: it is a science of how the world \emph{can} be.  

There are a number of different ways of understanding what this modality means. Perhaps the simplest in physics is to rely on the empirical fact that nature realises multiple instances of the same something in time and space.  All stones behave similarly when they fall and the same stone behaves similarly every time it falls.  This permits us to construct a space of possibilities and then use the regularities for predictions.  This structure can be seen as part of the elementary grammar of nature itself. And then the modality of physics and, consequently, the modality of the definition of meaning I have given are fully harmless against a serene and quite physicalism.  

But I nevertheless raise a small red flag here. Because we do not actually know the extent to which this structure is superimposed over the elementary texture of reality by ourselves. It could well be so: the structure could be generated 
precisely by the structure of the very `meaningful information' we have been concerned with here.  We are undoubtably limited parts of nature, and we are so even as understanders of this same nature.

\acknowledgements

I thank David Wolpert for private communications and especially Jenann Ismael for a critical reading of the article and very helpful suggestions. 

\begingroup\raggedright\endgroup


\end{document}